\documentclass[%
 reprint,
 amsmath,amssymb,
 aps,
]{revtex4-1}

\usepackage[pdftex]{graphicx}
\usepackage{dcolumn}
\usepackage{bm}

\begin{document}


\title{How to define the storage and loss moduli for a rheologically nonlinear material?}

\author{Ivan Argatov}
\email{ivan.argatov@campus.tu-berlin.de}
\affiliation{Institut f{\"u}r Mechanik, Technische Universit{\"a}t Berlin, 10623 Berlin, Germany}

\author{Alexei Iantchenko}
\affiliation{Faculty of Technology and Society, Malm\"o University, Malm\"o, Sweden
}
\author{Vitaly Kocherbitov}
\affiliation{Department of Biomedical Science, Faculty of Health and Society, Malm\"o University, Malm\"o, Sweden} 
\affiliation{Biofilms -- Research Center for Biointerfaces, Malm\"o University, Malm\"o, Sweden}

\date{\today}

\begin{abstract}
A large amplitude oscillatory shear (LAOS) is considered in the strain-controlled regime, and the interrelation between the Fourier transform (FT) and the stress decomposition (SD) approaches is established. Several definitions of the generalized storage and loss moduli are examined in a unified conceptual scheme based on the Lissajous--Bowditch plots. An illustrative example of evaluating the generalized moduli from a LAOS flow is given. 
\end{abstract}

\maketitle


\section{Introduction}
\label{secSL0}

Measuring rheological properties of complex fluids is usually performed in oscillatory shear flow by recording the time-dependent shear stress response to an externally applied oscillatory shear strain \citep{Larson1998}.

For a small strain amplitude $\gamma_0$, the measured shear stress is a simple harmonics with the same angular frequency $\omega$ and can be represented in the form
\begin{equation}
\sigma(t)=\gamma_0(G^\prime(\omega)\sin\omega t
+G^{\prime\prime}(\omega)\cos\omega t),
\label{1rh(0.1)}
\end{equation}
provided that the shear strain changes according to a sine law, i.e., $\gamma(t)=\gamma_0\sin\omega t$. The quantities $G^\prime(\omega)$ and $G^{\prime\prime}(\omega)$ are called the storage and loss moduli, respectively.

Equation (\ref{1rh(0.1)}) can be also represented in the form
\begin{equation}
\sigma(t)=\sigma_0\sin(\omega t+\delta),
\label{1rh(0.2)}
\end{equation}
where $\sigma_0=G_D(\omega)\gamma_0$ is the shear stress amplitude, $G_D(\omega)=\sqrt{G^\prime(\omega)^2+G^{\prime\prime}(\omega)^2}$ is the dynamic modulus. 

In many practical applications, monitoring changes of $G^\prime$ and $G^{\prime\prime}$ occurring in response to changes of environment variables is crucial for understanding the structure and dynamics of materials. For example,
the ratio $G^\prime/G^{\prime\prime}$ changes dramatically at the glass transition as a response to variation of relative humidity \citep{GrafKocherbitov2013,ZnamenskayaSotresGavryushov2013}. Since properties of the materials in the glassy and in the fluid state are totally different, they cannot be described by one model, and hence a model-free approach should be used for description rheology of materials in a broad range of values of temperature and relative humidity. Thus, calculation of $G^\prime$ and $G^{\prime\prime}$ from both traditional rheometry data and from newer methods such as QCM-D 
\citep{GrafKocherbitov2013,BjorklundKocherbitov2015} finds many applications in pharmacy, biophysics, surface and colloid chemistry.

For Small Amplitude Oscillatory Shear (SAOS)  flows, the amplitude ratio, $\sigma_0/\gamma_0$,  and the loss angle, $\delta(\omega)$, are independent of $\gamma_0$, whereas the moduli $G^\prime(\omega)$ and $G^{\prime\prime}(\omega)$ represent elastic and viscous material properties. The quantities $G^\prime(\omega)$ and $G^{\prime\prime}(\omega)$ represent integral characteristics of the material functions (see, e.g., \citep{Christensen1971,Pipkin1986,BraderSiebenburgerBallauff2010}), and in SAOS they bear complete information on viscoelastic properties. Recently, the so-called incomplete storage and loss moduli were introduced in \citep{Argatov2012} to describe sinusoidally driven testing on a finite interval of time. 

However, when the assumption of small strain amplitudes is not satisfied, the same sinusoidal excitation will elicit a non-sinusoidal shear stress response, which is called Large Amplitude Oscillatory Shear (LAOS) flow. While in SAOS, the storage and loss moduli possess clear physical meanings, these parameters lose their physical significance in the nonlinear regime \citep{ChoHyunAhn2005}.

There is still an urgent need for finding measurable and physically meaningful material parameters that can conquer the status of nonlinear counterparts of $G^\prime(\omega)$ and $G^{\prime\prime}(\omega)$ \citep{ThompsonAlickedeSouza2015}.
In the present paper, we consider the problem from two viewpoints: Fourier transform (FT) and stress decomposition (SD). Correspondingly, the FT and SD coefficients are denoted by $G^\prime_m$, $G^{\prime\prime}_m$ and ${\bf G}^\prime_m$, ${\bf G}^{\prime\prime}_m$, $m=1,3,\ldots$, while the generalized storage and loss moduli are indicated using letter subscripts. 

\section{Shear strain-control LAOS: Relation between the FT and SD approaches}
\label{secSL1}

Let us assume that the shear strain input is represented as a sine wave
\begin{equation}
\gamma(t)=\gamma_0\sin\omega t,
\label{1rh(1.1)}
\end{equation}
so that the strain rate is given by 
\begin{equation}
\dot\gamma(t)=\omega\gamma_0\cos\omega t.
\label{1rh(1.2)}
\end{equation}

In the framework of the Fourier-Transform (FT) rheology, the nonlinear stress response is represented in the form of a Fourier series
\begin{equation}
\sigma=\gamma_0\sum_{m:{\rm odd}}
G^\prime_m(\omega,\gamma_0)\sin m\omega t
+G^{\prime\prime}_m(\omega,\gamma_0)\cos m\omega t,
\label{1rh(1.3)}
\end{equation}
where the Fourier-coefficient moduli $G^\prime_m(\omega,\gamma_0)$ and $G^{\prime\prime}_m(\omega,\gamma_0)$ are, in general, functions of angular frequency $\omega$ and strain amplitude $\gamma_0$, which are the two LAOStrain input parameters. 

In the framework of the Stress Decomposition (SD) approach of \citet{ChoHyunAhn2005}, the stress response can be described as
\begin{equation}
\sigma=\sum_{m:{\rm odd}}
{\bf G}^\prime_m(\omega,\gamma_0)\gamma^m
+{\bf G}^{\prime\prime}_m(\omega,\gamma_0)\Bigl(
\frac{\dot\gamma}{\omega}\Bigr)^m,
\label{1rh(1.4)}
\end{equation}
so that the stress response represents the sum
\begin{equation}
\sigma=\sigma^\prime(\gamma,\gamma_0)
+\sigma^{\prime\prime}(\dot\gamma,\gamma_0)
\label{1rh(1.5)}
\end{equation}
of the elastic and viscous contributions
\begin{equation}
\sigma^\prime(\gamma,\gamma_0)=\sum_{n=0}^\infty
{\bf G}^\prime_{2n+1}(\omega,\gamma_0)\gamma^{2n+1},
\label{1rh(1.6)}
\end{equation}
\begin{equation}
\sigma^{\prime\prime}(\dot\gamma,\gamma_0)=\sum_{n=0}^\infty
{\bf G}^{\prime\prime}_{2n+1}(\omega,\gamma_0)
\frac{\dot\gamma^{2n+1}}{\omega^{2n+1}}.
\label{1rh(1.7)}
\end{equation}

It is well known \citep{EwoldtHosoiMcKinley2008,YuWangZhou2009,Ewoldt2013} that the two forms (\ref{1rh(1.3)}) and (\ref{1rh(1.4)}) are equivalent, and the Fourier-coefficient moduli $G^\prime_m$ and $G^{\prime\prime}_m$ can be expressed in terms of the power-coefficient moduli ${\bf G}^\prime_{2n+1}$ and ${\bf G}^{\prime\prime}_{2n+1}$ and vise versa. 

\subsection{FT coefficients via SD coefficients}
\label{secSL1A}

First, making use of the trigonometric formulas
$$
\sin^{2n+1}x=\sum_{k=0}^n
\frac{(-1)^{n+k}}{2^{2n}}
\biggl({2n+1\atop k}\biggr)\sin(2n-2k+1)x,
$$
$$
\cos^{2n+1}x=\frac{1}{2^{2n}}\sum_{k=0}^n
\biggl({2n+1\atop k}\biggr)\cos(2n-2k+1)x,
$$
(see, e.g., formulas (1.320.3) and (1.320.7) given by \citet{GradshteynRyzhik2000}), we find
\begin{equation}
G^\prime_m=(-1)^{(m-1)/2}\sum_{n=(m-1)/2}^\infty
\alpha_{n,m}
{\bf G}^\prime_{2n+1}\gamma^{2n}_0,
\label{1rh(1.8)}
\end{equation}
\begin{equation}
G^{\prime\prime}_m=\sum_{n=(m-1)/2}^\infty
\alpha_{n,m}
{\bf G}^{\prime\prime}_{2n+1}\gamma^{2n}_0,
\label{1rh(1.9)}
\end{equation}
where the coefficients 
$$
\alpha_{n,m}=\frac{1}{2^{2n}}\biggl({2n+1\atop (2n-m+1)/2}\biggr)
$$
are expressed in terms of a binomial coefficient
$$
\Bigl({p\atop k}\Bigr)=\frac{p!}{k!(p-k)!}, \quad
\Bigl({p\atop 0}\Bigr)=1,
$$
with $p!=1\cdot 2\cdot \ldots(p-1)p$ being the factorial of $p$.

Observe that the moduli $G^\prime_m$ , $G^{\prime\prime}_m$ and ${\bf G}^\prime_m$, ${\bf G}^{\prime\prime}_m$ have the same dimension of stress. 

\subsection{SD coefficients via FT coefficients}
\label{secSL1B}

Now, we make use of the following trigonometric formulas (see, e.g., formulas (1.332.2) and (1.331.3) in \citep{GradshteynRyzhik2000}):
$$
\sin mx=\sum_{n=0}^{(m-1)/2} S_{m,2n+1}\sin^{2n+1}x,\quad
S_{m,1}=m,
$$
$$
S_{m,p}=(-1)^{(p-1)/2}\frac{m}{p!}\prod_{k=0}^{(p-3)/2}[m^2-(2k+1)^2],\quad p\not=1,
$$
$$
\cos mx=\sum_{n=0}^{(m-1)/2} C_{m,2n+1}\cos^{2n+1}x,
$$
$$
C_{m,p}=\frac{(-1)^{(m-p)/2}m 2^p}{m-p}
\biggl({(m+p-2)/2 \atop (m-p-2)/2}\biggr),
$$
$$
p=1,3,\ldots,m-2,\quad C_{m,m}=2^{m-1}.
$$
Here, $m$ is assumed to be odd. 

In this way, we arrive at the relations
\begin{equation}
{\bf G}^\prime_m=\frac{1}{\gamma_0^{m-1}}\sum_{n=(m-1)/2}^\infty
G^\prime_{2n+1} S_{2n+1,m},
\label{1rh(1.10)}
\end{equation}
\begin{equation}
{\bf G}^{\prime\prime}_m=\frac{1}{\gamma_0^{m-1}}\sum_{n=(m-1)/2}^\infty
G^{\prime\prime}_{2n+1} C_{2n+1,m}.
\label{1rh(1.11)}
\end{equation}

In particular, we find that 
\begin{equation}
{\bf G}^\prime_1=G^\prime_1+\sum_{n=1}^\infty
(2n+1)G^\prime_{2n+1},
\label{1rh(1.12)}
\end{equation}
\begin{equation}
{\bf G}^{\prime\prime}_1=G^{\prime\prime}_1+\sum_{n=1}^\infty
(-1)^n(2n+1) G^{\prime\prime}_{2n+1}.
\label{1rh(1.13)}
\end{equation}

Finally, we underline that if the Fourier series representation (\ref{1rh(1.3)}) contains a finite number of terms, that is $m=1,3,\ldots,M$, where $M$ is odd, then the stress decomposition (\ref{1rh(1.4)}) will contain terms up to the degree $M$, including. This can be seen from formulas (\ref{1rh(1.10)}) and (\ref{1rh(1.11)}), where the summation starts with $(m-1)/2$. 

\section{Generalized moduli for LAOS}
\label{secSL2}

\subsection{Introducing the storage and loss moduli via the Krylov--Bogoliubov equivalent linearization}
\label{secSL2A}

In their studies of quasilinear oscillations of one-degree-of-freedom systems, 
\citet{KrylovBogoliubov1947} approximately replaced a nonlinear function $\sigma(\gamma,\dot\gamma)$ by a linear function of $\gamma$ and $\dot\gamma$ with coefficients depending on the amplitude and frequency of oscillations, that is
\begin{equation}
\sigma(\gamma,\dot\gamma)\simeq G^\prime_E(\omega,\gamma_0)\gamma
+\frac{G^{\prime\prime}_E(\omega,\gamma_0)}{\omega}\dot\gamma,
\label{1rh(2.1)}
\end{equation}
where the so-called equivalent storage $G^\prime_E$ and loss $G^{\prime\prime}$ moduli are given by
\begin{equation}
G^\prime_E=\frac{1}{\pi\gamma_0}\int\limits_0^{2\pi}
\sigma(\gamma_0\cos\psi,-\omega\gamma_0\sin\psi)\cos\psi\,d\psi,
\label{1rh(2.2)}
\end{equation}
\begin{equation}
G^{\prime\prime}_E=-\frac{1}{\pi\gamma_0}\int\limits_0^{2\pi}
\sigma(\gamma_0\cos\psi,-\omega\gamma_0\sin\psi)\sin\psi\,d\psi.
\label{1rh(2.3)}
\end{equation}

Let us rewrite the Krylov--Bogoliubov formulas in order to facilitate the comparison with another results. First, utilizing the change of the integration variable $\psi=\tilde\psi-\pi/2$, we can recast formulas (\ref{1rh(2.2)}) and (\ref{1rh(2.3)}) as
\begin{equation}
G^\prime_E=\frac{1}{\pi\gamma_0}\int\limits_0^{2\pi}
\sigma(\gamma_0\sin\psi,\omega\gamma_0\cos\psi)\sin\psi\,d\psi,
\label{1rh(2.4)}
\end{equation}
\begin{equation}
G^{\prime\prime}_E=\frac{1}{\pi\gamma_0}\int\limits_0^{2\pi}
\sigma(\gamma_0\sin\psi,\omega\gamma_0\cos\psi)\cos\psi\,d\psi.
\label{1rh(2.5)}
\end{equation}

Now, recalling that $\gamma=\gamma_0\sin\omega t$ and $\dot\gamma=\omega\gamma_0\cos\omega t$ (see Eqs.~(\ref{1rh(1.1)}) and (\ref{1rh(1.2)})), we can replace the definite integrals in (\ref{1rh(2.4)}) and (\ref{1rh(2.5)}) with contour integrals as follows:
\begin{equation}
G^\prime_E=-\frac{1}{\pi\omega\gamma_0^2}\oint
\sigma(\gamma,\dot\gamma)\,d\dot\gamma,
\label{1rh(2.6)}
\end{equation}
\begin{equation}
G^{\prime\prime}_E=\frac{1}{\pi\gamma_0^2}\oint
\sigma(\gamma,\dot\gamma)\,d\gamma.
\label{1rh(2.7)}
\end{equation}

Thus, it can be seen that the equivalent moduli $G^\prime_E$ and $G^{\prime\prime}_E$ {\it per se\/} correspond to the generalized dynamic moduli introduced by \citet{ChoHyunAhn2005}. The energy and geometric interpretations of these moduli were given by \citet{IlyinKulichikhinMalkin2014}. 

The substitution of the FT expansion (\ref{1rh(1.3)}), which, by introducing the phase angle $\psi=\omega t$, van be rewritten in the form
$$
\sigma=\gamma_0\sum_{m:{\rm odd}}
G^\prime_m(\omega,\gamma_0)\sin m\psi
+G^{\prime\prime}_m(\omega,\gamma_0)\cos m\psi,
$$
into formulas (\ref{1rh(2.4)}) and (\ref{1rh(2.5)}) yields 
\begin{equation}
G^\prime_E(\omega,\gamma_0)=G^\prime_1(\omega,\gamma_0),
\label{1rh(2.8a)}
\end{equation}
\begin{equation}
G^{\prime\prime}_E(\omega,\gamma_0)=G^{\prime\prime}_1(\omega,\gamma_0).
\label{1rh(2.8b)}
\end{equation}

On the other hand, for the SD expansion (\ref{1rh(1.4)}), the following relations hold \citep{ChoHyunAhn2005}:
\begin{equation}
G^\prime_E(\omega,\gamma_0)={\bf G}^\prime_1(\omega)
+\sum_{n=1}^\infty
\beta_{2n+1}{\bf G}^\prime_{2n+1}(\omega,\gamma_0)\gamma^{2n}_0,
\label{1rh(2.9)}
\end{equation}
\begin{equation}
G^{\prime\prime}_E(\omega,\gamma_0)={\bf G}^{\prime\prime}_1(\omega)
+\sum_{n=1}^\infty
\beta_{2n+1}{\bf G}^{\prime\prime}_{2n+1}(\omega,\gamma_0)
\gamma^{2n}_0.
\label{1rh(2.10)}
\end{equation}
Here we have introduced the notation
$$
\beta_{2n+1}=\frac{4}{\pi}\int\limits_0^{\pi/2}
\cos^{2n+2}\psi\,d\psi=\frac{(2n+2)!}{2^{2n+1}[(n+1)!]^2}.
$$

Note that formulas (\ref{1rh(2.8a)})--(\ref{1rh(2.10)}) agree with formulas (\ref{1rh(1.9)}) and (\ref{1rh(1.8)}) for $m=1$.

\subsection{Minimum and large strain moduli}
\label{secSL2B}

For LAOStrain experiments, \citet{EwoldtHosoiMcKinley2008} introduced the minimum-strain modulus, $G^\prime_M$, and the large-strain modulus, $G^\prime_L$, defined as
\begin{equation}
G^\prime_M=\frac{d\sigma}{d\gamma}\Bigr\vert_{\gamma=0},\quad
G^\prime_L=\frac{\sigma}{\gamma}\Bigr\vert_{\gamma=\pm\gamma_0}.
\label{1rh(3.1)}
\end{equation}

Taking into account that
\begin{equation}
\frac{d\sigma}{d\gamma}=\frac{d\sigma}{dt}\frac{dt}{d\gamma}
=\frac{d\sigma}{dt}\frac{1}{(d\gamma/dt)}
=\frac{\dot\sigma}{\dot\gamma},
\label{1rh(3.2)}
\end{equation}
the minimum-strain and large-strain moduli (\ref{1rh(3.1)}) can be evaluated in terms of the FT-coefficient moduli as follows \citep{EwoldtHosoiMcKinley2008,LaugerStettin2010}:
\begin{equation}
G^\prime_M=\sum_{m:{\rm odd}}
m G^\prime_m(\omega,\gamma_0),
\label{1rh(3.3)}
\end{equation}
\begin{equation}
G^\prime_L=\sum_{m:{\rm odd}}
(-1)^{(m-1)/2} G^\prime_m(\omega,\gamma_0).
\label{1rh(3.4)}
\end{equation}

In the framework of the SD approach, in light of (\ref{1rh(1.5)})--(\ref{1rh(1.7)}) and (\ref{1rh(3.2)}), we will have
\begin{eqnarray}
\frac{d\sigma}{d\gamma} & = & {\bf G}^\prime_1(\omega)
+\sum_{n=1}^\infty
(2n+1){\bf G}^\prime_{2n+1}(\omega,\gamma_0)\gamma^{2n}
\nonumber\\
{} & {} & {}+\sum_{n=0}^\infty
(2n+1){\bf G}^{\prime\prime}_{2n+1}(\omega,\gamma_0)
\frac{\dot\gamma^{2n-1}\ddot\gamma}{\omega^{2n+1}},
\nonumber
\end{eqnarray}
and, therefore, taking into account that $\ddot\gamma\bigr\vert_{\gamma=0}=0$, we obtain 
\begin{equation}
G^\prime_M={\bf G}^\prime_1(\omega).
\label{1rh(3.5)}
\end{equation}

Further, the substitution of (\ref{1rh(1.4)}) into the second formula (\ref{1rh(3.1)}) yields 
\begin{equation}
G^\prime_L={\bf G}^\prime_1(\omega)
+\sum_{n=1}^\infty
{\bf G}^\prime_{2n+1}(\omega,\gamma_0)\gamma^{2n}.
\label{1rh(3.6)}
\end{equation}
It is important to note \citep{HessAksel2011} that the moduli $G^\prime_M$ and $G^\prime_L$ can be evaluated via two different methods: by stress decomposition, e.g., using FT rheology, or from Lissajous--Bowditch plots.

Similarly, \citep{EwoldtHosoiMcKinley2008} introduced the minimum-rate dynamic viscosity, $\eta^\prime_M$, and  the large-rate dynamic viscosity, $\eta^\prime_L$, defined as
\begin{equation}
\eta^\prime_M=\frac{d\sigma}{d\dot\gamma}\Bigr\vert_{\dot\gamma=0},\quad
\eta^\prime_L=\frac{\sigma}{\dot\gamma}\Bigr\vert_{\dot\gamma=\pm\dot\gamma_0},
\label{1rh(3.7)}
\end{equation}
where $\dot\gamma_0=\omega\gamma_0$.

In the framework of the FT rheology, the following expansions take place \citep{EwoldtHosoiMcKinley2008,LaugerStettin2010}:
\begin{equation}
\eta^\prime_M=\frac{1}{\omega}\sum_{m:{\rm odd}}
(-1)^{(m-1)/2}m G^{\prime\prime}_m(\omega,\gamma_0),
\label{1rh(3.8)}
\end{equation}
\begin{equation}
\eta^\prime_L=\frac{1}{\omega}\sum_{m:{\rm odd}}
G^{\prime\prime}_m(\omega,\gamma_0).
\label{1rh(3.9)}
\end{equation}

In the framework of the SD rheology, the substitution of (\ref{1rh(1.4)}) into the second formula (\ref{1rh(3.4)}) yields 
\begin{equation}
\eta^\prime_L=\frac{1}{\omega}\biggl(
{\bf G}^{\prime\prime}_1(\omega)
+\sum_{n=1}^\infty
{\bf G}^{\prime\prime}_{2n+1}(\omega,\gamma_0)\gamma_0^{2n}\biggl).
\label{1rh(3.10)}
\end{equation}

To evaluate the dynamic viscosity $\eta^\prime_M$, we note that
$$
\frac{d\sigma}{d\dot\gamma}=\frac{d\sigma}{dt}\frac{dt}{d\dot\gamma}
=\frac{\dot\sigma}{\ddot\gamma},
$$
where, in light of (\ref{1rh(1.1)}) and (\ref{1rh(1.2)}), $\ddot\gamma=-\omega^2\gamma_0\sin\omega t$, so that $\ddot\gamma=-\omega^2\gamma$ and, therefore, 
\begin{eqnarray}
\frac{d\sigma}{d\dot\gamma} & = & -\frac{\dot\gamma}{\omega^2\gamma}\biggl(
{\bf G}^\prime_1(\omega)
+\sum_{n=1}^\infty
(2n+1){\bf G}^\prime_{2n+1}(\omega,\gamma_0)\gamma^{2n}\biggr)
\nonumber\\
{} & {} & {}+\frac{1}{\omega}{\bf G}^{\prime\prime}_1(\omega)
+\sum_{n=1}^\infty
(2n+1){\bf G}^{\prime\prime}_{2n+1}(\omega,\gamma_0)
\frac{\dot\gamma^{2n}}{\omega^{2n-1}}.
\nonumber
\end{eqnarray}

Thus, the minimum-rate dynamic viscosity is given by
\begin{equation}
\eta^\prime_M=\frac{1}{\omega}{\bf G}^{\prime\prime}_1(\omega).
\label{1rh(3.11)}
\end{equation}

Observe that the large-strain modulus $G^\prime_L$, as it is given by the power series expansion (\ref{1rh(3.6)}), coincides with one of the generalized storage moduli introduced by \citet{ChoHyunAhn2005} (see also \citep{ChoSongChang2010}) as follows:
\begin{equation}
G^\prime_L=\frac{\sigma^\prime_{\rm max}}{\gamma_0},\quad
G^{\prime\prime}_L=\frac{\sigma^{\prime\prime}_{\rm max}}{\gamma_0}.
\label{1rh(3.12)}
\end{equation}
Here, $\sigma^\prime_{\rm max}$ and $\sigma^{\prime\prime}_{\rm max}$ are the maxima of the elastic (\ref{1rh(1.6)}) and viscous (\ref{1rh(1.7)}) stresses, which are attained at $\gamma=\gamma_0$ and $\dot\gamma=\dot\gamma_0$, respectively. 

Further, observe that the generalized loss modulus $G^{\prime\prime}_L$ is equal to $\omega\eta^\prime_L$, that is
\begin{equation}
G^{\prime\prime}_L=\omega\eta^\prime_L=\omega
\frac{\sigma}{\dot\gamma}\Bigr\vert_{\dot\gamma=\pm\dot\gamma_0}.
\label{1rh(3.13)}
\end{equation}

By analogy, we can introduce the generalized minimum-rate loss modulus as
\begin{equation}
G^{\prime\prime}_M=\omega\eta^\prime_M=\omega
\frac{d\sigma}{d\dot\gamma}\Bigr\vert_{\dot\gamma=0}.
\label{1rh(3.14)}
\end{equation}

Therefore, in view of (\ref{1rh(3.9)}), (\ref{1rh(3.10)}), and (\ref{1rh(3.13)}), we will have
\begin{eqnarray}
G^{\prime\prime}_L & = & \sum_{m:{\rm odd}}
G^{\prime\prime}_m(\omega,\gamma_0)
\label{1rh(3.15)}\\
{} & = & {\bf G}^{\prime\prime}_1(\omega)
+\sum_{n=1}^\infty
{\bf G}^{\prime\prime}_{2n+1}(\omega,\gamma_0)\gamma_0^{2n}.
\label{1rh(3.16)}
\end{eqnarray}

Correspondingly, formulas (\ref{1rh(3.8)}), (\ref{1rh(3.11)}), and (\ref{1rh(3.14)}) give
\begin{eqnarray}
G^{\prime\prime}_M & = & \sum_{m:{\rm odd}}
(-1)^{(m-1)/2}m G^{\prime\prime}_m(\omega,\gamma_0)
\label{1rh(3.17)}\\
{} & = & {\bf G}^{\prime\prime}_1(\omega).
\label{1rh(3.18)}
\end{eqnarray}

It is interesting to note that the generalized storage and loss moduli $G^\prime_M$ and $G^{\prime\prime}_M$ coincide with the first SD coefficients ${\bf G}^\prime_1$ and ${\bf G}^{\prime\prime}_1$.

\subsection{Minimum and large stress moduli}
\label{secSL2C}

Now, let us introduce the minimum-stress modulus, ${\cal G}^\prime_m$, and the large-stress modulus, ${\cal G}^\prime_l$, defined as follows:
\begin{equation}
{\cal G}^\prime_m=\frac{d\sigma}{d\gamma}\Bigr\vert_{\sigma=0},\quad
{\cal G}^\prime_l=\frac{\sigma}{\gamma}\Bigr\vert_{\sigma=\pm\sigma_0}.
\label{1rh(5.1)}
\end{equation}
Here, $\sigma_0$ is the stress amplitude, i.e., $\sigma_0=\max\sigma(t)$. Note that the quantity ${\cal G}^\prime_m$ was originally introduced by \citet{RogersErwinVlassopoulos2011} as the so-called apparent cage modulus. 

Similarly, let us define the minimum-stress dynamic viscosity, $\eta^\prime_m$, and  the large-stress dynamic viscosity, $\eta^\prime_l$, defined by the following formulas (cf. (\ref{1rh(3.7)})):
\begin{equation}
\eta^\prime_m=\frac{d\sigma}{d\dot\gamma}\Bigr\vert_{\sigma=0},\quad
\eta^\prime_l=\frac{\sigma}{\dot\gamma}\Bigr\vert_{\sigma=\pm\sigma_0}.
\label{1rh(5.2)}
\end{equation}
where $\dot\gamma_0=\omega\gamma_0$.

\begin{figure}[h]
\centering
\includegraphics[scale=0.3]{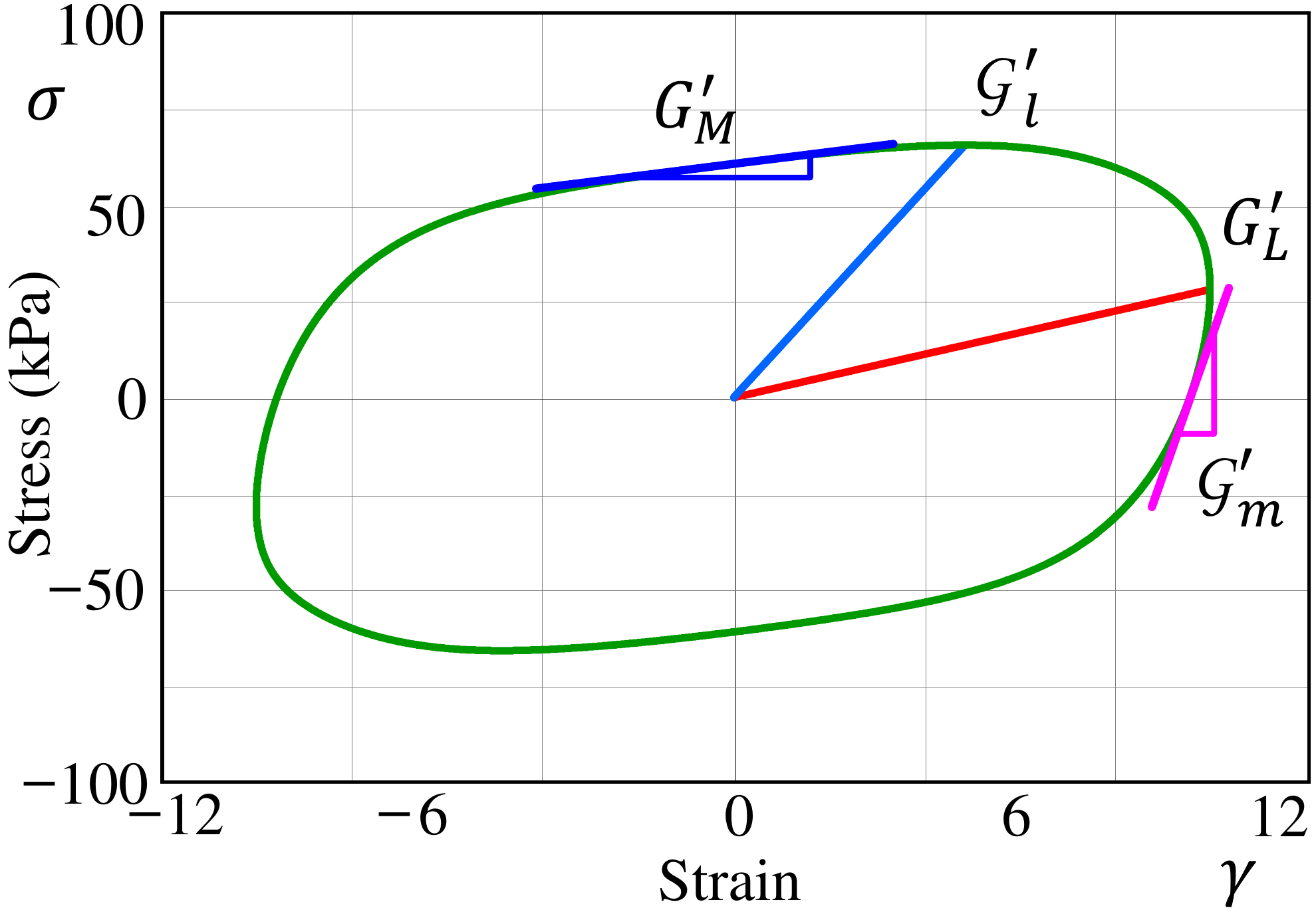}
\caption{Lissajous--Bowditch plot of stress versus strain. Four straight lines represent the defined moduli for LAOS measurements under a sinusoidal strain input.}
\label{fig:Fig4a}
\end{figure}

\begin{figure}[h]
\centering
\includegraphics[scale=0.3]{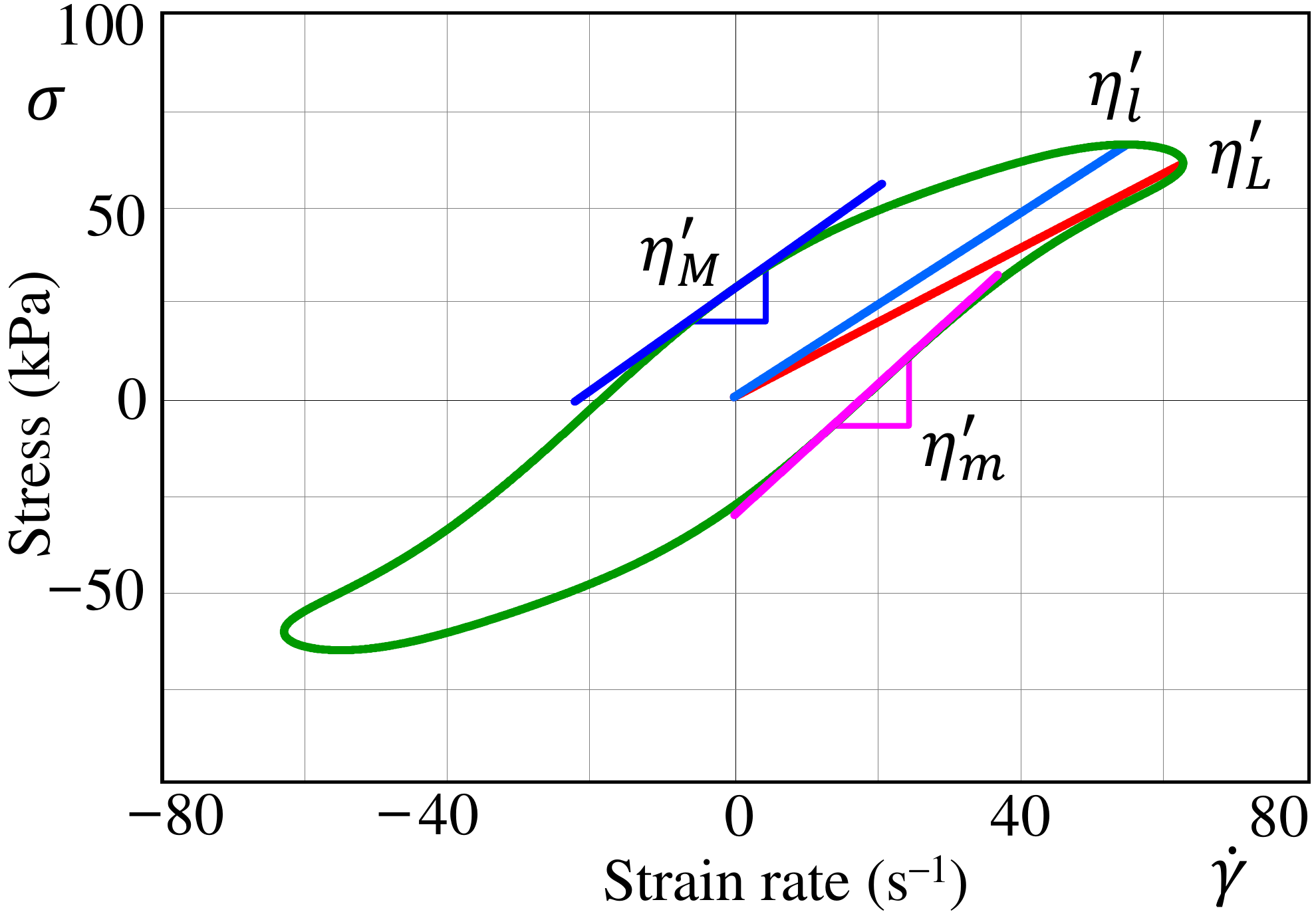}
\caption{Lissajous--Bowditch plot of stress versus strain rate. Four straight lines represent the defined viscosities for LAOS measurements under a sinusoidal strain input.}
\label{fig:Fig4b}
\end{figure}

In Figs.~\ref{fig:Fig4a} and \ref{fig:Fig4b}, an illustrative example (see Section~\ref{secSL6A}) of Lissajous--Bowditch plots for a complete cycle with a sinusoidal strain input is represented together straight lines corresponding to the introduced above moduli and viscosities. 

In the framework of SAOS, it is easy to get interpretations for the moduli (\ref{1rh(5.1)}) and viscosities (\ref{1rh(5.2)}). Indeed, if the applied oscillatory strain is given by $\gamma(t)=\gamma_0\sin\omega t$, then the oscillatory stress response can be represented in the form $\sigma(t)=\sigma_0\sin(\omega t+\delta)$. Then, it can be shown that
\begin{equation}
{\cal G}^\prime_m={\cal G}^\prime_l=\frac{\sigma_0}{\gamma_0\cos\delta},\quad
\eta^\prime_m=\eta^\prime_l=\frac{\sigma_0}{\gamma_0\sin\delta}.
\label{1rh(5.3)}
\end{equation}

Therefore, taking into account the SAOS relations
$$
G_D=\frac{\sigma_0}{\gamma_0},\quad
G_D^2=G^{\prime 2}+G^{\prime\prime 2},
$$
$$
G^\prime=G_D\cos\delta,\quad G^{\prime\prime}=G_D\sin\delta,
$$
where $G^\prime(\omega)$ and $G^{\prime\prime}(\omega)$ are the storage and loss moduli, we readily get
\begin{equation}
{\cal G}^\prime_m={\cal G}^\prime_l=\frac{G_D}{\cos\delta}=\frac{G_D^2}{G^\prime}
\label{1rh(5.4)}
\end{equation}
and
\begin{equation}
\eta^\prime_m=\eta^\prime_l=\frac{G_D}{\omega\sin\delta}
=\frac{G_D^2}{\omega G^{\prime\prime}}.
\label{1rh(5.5)}
\end{equation}

Note that the first formula (\ref{1rh(5.3)}) and (\ref{1rh(5.4)}) were obtained for the cage modulus ${\cal G}^\prime_m$ by \citet{RogersErwinVlassopoulos2011}.

Furthermore, in the framework of SAOS, the following inverse relations take place:
\begin{equation}
G^\prime+i G^{\prime\prime}=\frac{1}{J^\prime-i J^{\prime\prime}},\quad
J^\prime=\frac{G^\prime}{G_D^2},\quad
J^{\prime\prime}=\frac{G^{\prime\prime}}{G_D^2}.
\label{1rh(5.6)}
\end{equation}
Here, $J^\prime(\omega)$ and $J^{\prime\prime}(\omega)$ are the storage and loss compliances, respectively.

Thus, from (\ref{1rh(5.4)})--(\ref{1rh(5.6)}), it follows that
\begin{equation}
{\cal G}^\prime_m={\cal G}^\prime_l=\frac{1}{J^\prime},\quad
\eta^\prime_m=\eta^\prime_l=\frac{1}{\omega J^{\prime\prime}}.
\label{1rh(5.7)}
\end{equation}

We emphasize that formulas (\ref{1rh(5.7)}) hold only in the SAOS regime. In the case of LAOS, by analogy with (\ref{1rh(5.7)}), one can introduce the following quantities:
\begin{equation}
J^\prime_m(\omega,\gamma_0)=\frac{1}{{\cal G}^\prime_m(\omega,\gamma_0)}
=\frac{d\gamma}{d\sigma}\Bigr\vert_{\sigma=0},
\label{1rh(5.8a)}
\end{equation}
\begin{equation}
J^\prime_l(\omega,\gamma_0)=\frac{1}{{\cal G}^\prime_l(\omega,\gamma_0)}
=\frac{\gamma}{\sigma}\Bigr\vert_{\sigma=\pm\sigma_0},
\label{1rh(5.8b)}
\end{equation}
\begin{equation}
J^{\prime\prime}_m(\omega,\gamma_0)=\frac{1}{\omega\eta^\prime_m(\omega,\gamma_0)}
=\frac{1}{\omega}\frac{d\dot\gamma}{d\sigma}\Bigr\vert_{\sigma=0},
\label{1rh(5.9a)}
\end{equation}
\begin{equation}
J^{\prime\prime}_l(\omega,\gamma_0)=\frac{1}{\omega\eta^\prime_l(\omega,\gamma_0)}
=\frac{1}{\omega}\frac{\dot\gamma}{\sigma}\Bigr\vert_{\sigma=\pm\sigma_0}.
\label{1rh(5.9b)}
\end{equation}

\begin{figure}[b]
\centering
\includegraphics[scale=0.3]{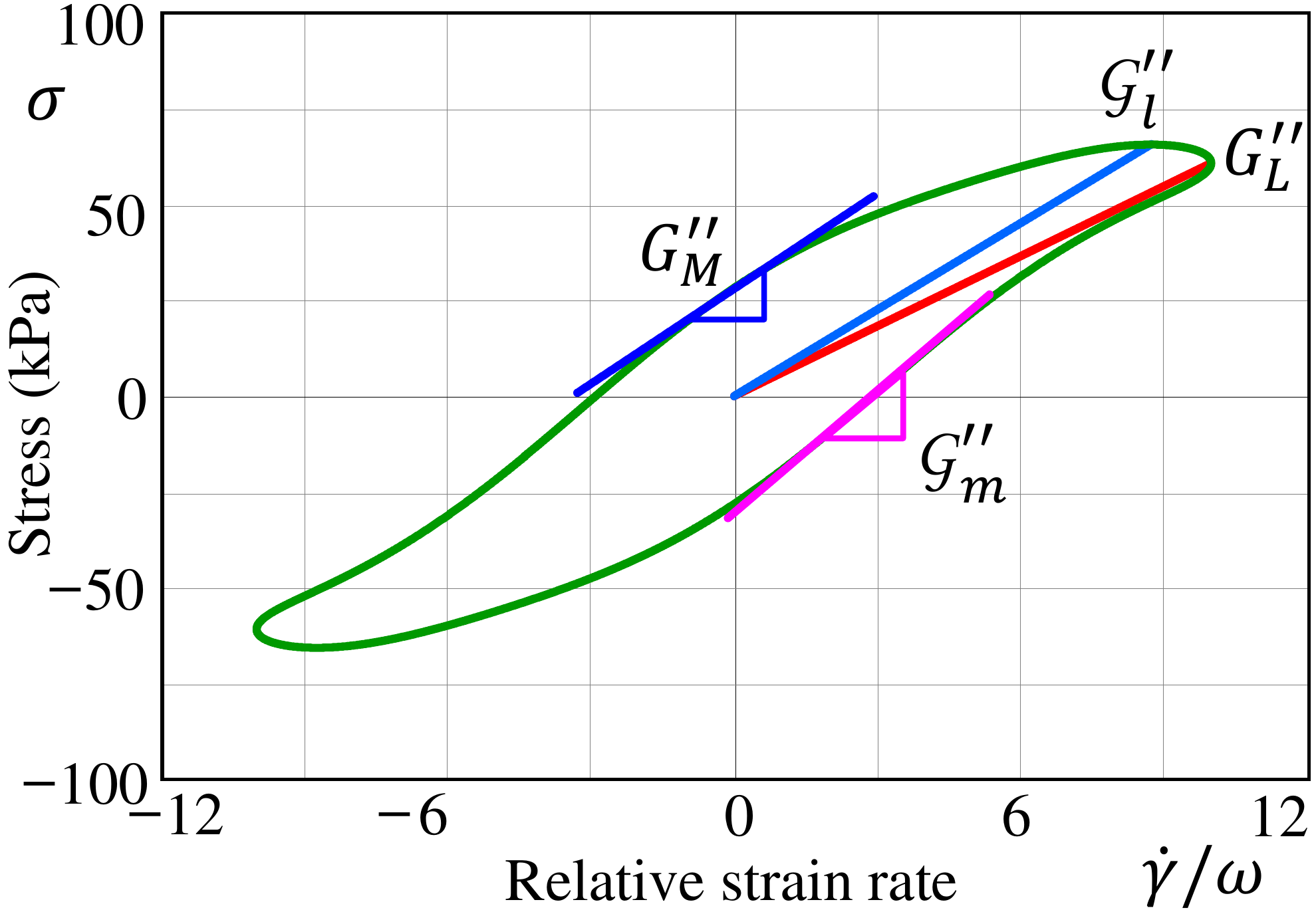}
\caption{Lissajous--Bowditch plot of stress versus relative strain rate. Four straight lines represent the defined moduli for LAOS measurements under a sinusoidal strain input.}
\label{fig:Fig6}
\end{figure}

The quantities (\ref{1rh(5.8a)})--(\ref{1rh(5.9b)}) can be called generalized storage and loss compliances defined in the strain-controlled test. 

By analogy with Eqs.~(\ref{1rh(5.7)})--(\ref{1rh(5.8b)}), we can introduce the following quantities:
\begin{equation}
{\cal G}^{\prime\prime}_m(\omega,\gamma_0)
=\frac{1}{J^{\prime\prime}_m(\omega,\gamma_0)},
\label{1rh(5.10a)}
\end{equation}
\begin{equation}
{\cal G}^{\prime\prime}_l(\omega,\gamma_0)
=\frac{1}{J^{\prime\prime}_l(\omega,\gamma_0)}.
\label{1rh(5.10b)}
\end{equation}

Note that, though the moduli ${\cal G}^{\prime\prime}_m$ and ${\cal G}^{\prime\prime}_l$ were introduced via the corresponding viscosities (see formulas (\ref{1rh(5.9a)})--(\ref{1rh(5.10b)})), they can be directly measured from the scaled Lissajous--Bowditch plot (see Fig.~\ref{fig:Fig6}).

\subsection{Generalization of the dynamic modulus and the loss angle}
\label{secSL2D}

In LAOS, as it was observed by \citet{Rogers2012}, it is still possible to define a value of the dynamic modulus
\begin{equation}
G_D(\omega,\gamma_0)=\frac{\sigma_0}{\gamma_0},
\label{1rh(4.1)}
\end{equation}
where $\sigma_0={\rm max\,}\sigma(t)$ is the stress amplitude. 

\begin{figure}[h]
\centering
\includegraphics[scale=0.3]{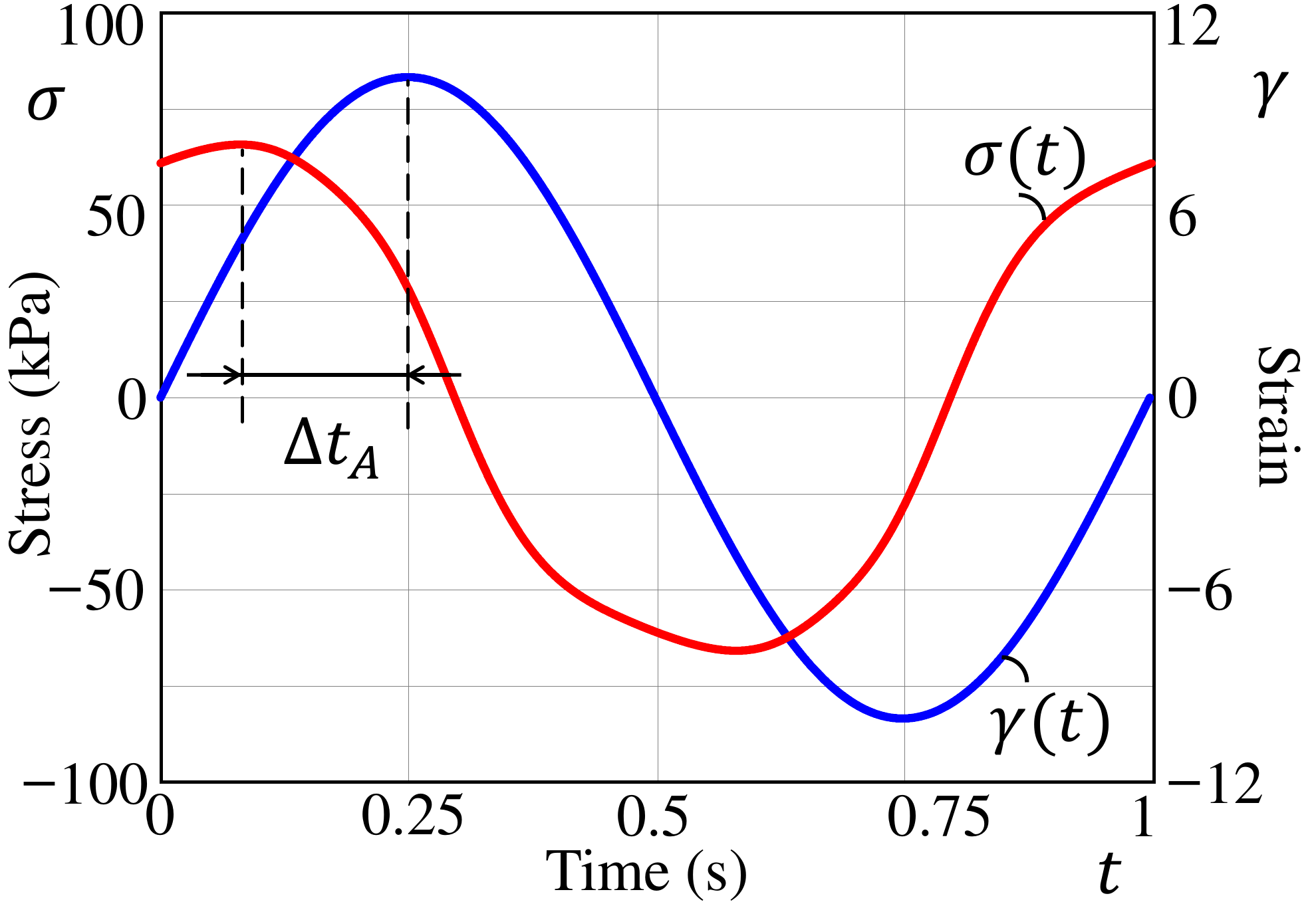}
\caption{Sinusoidal strain input and non-dimensional stress output.}
\label{fig:Fig3}
\end{figure}

The generalized loss angle can be introduced in a number of ways. First, based on the time delay, ${\Delta t}_A$, between the amplitude values of the strain input and the stress output, i.e., between the stress and strain maxima (see Fig.~\ref{fig:Fig3}), we can introduce the so-called amplitude based loss angle
\begin{equation}
\delta_A(\omega,\gamma_0)=\omega{\Delta t}_A.
\label{1rh(4.2)}
\end{equation}

The corresponding storage and loss moduli can be formally introduced by the formulas
\begin{equation}
G_A^\prime=G_D\cos\delta_A, \quad
G_A^{\prime\prime}=G_D\sin\delta_A,
\label{1rh(4.3)}
\end{equation}
where the dynamic modulus $G_D$ is given by (\ref{1rh(4.1)}).

Further, as it was noted by \citet{RogersErwinVlassopoulos2011}, in the nonlinear regime, the concept of a single valued phase angle only makes sense when the periods of oscillation are identical (equal to $2\pi/\omega$) and when the phase difference is measured at zero stress. In this case, the so-called zero-stress based generalized loss angle, $\delta_{Z\sigma}$, can be introduced via the following formula \citet{RogersErwinVlassopoulos2011}:
\begin{equation}
\delta_{Z\sigma}(\omega,\gamma_0)={\rm arcsin\,}\frac{\gamma_{Z\sigma}}{\gamma_0}.
\label{1rh(4.4)}
\end{equation}
Here, $\gamma_{Z\sigma}$ is the strain at which the stress is instantaneously zero. This quantity can be evaluated from the Lissajous--Bowditch curve. 

\begin{figure}[h]
\centering
\includegraphics[scale=0.3]{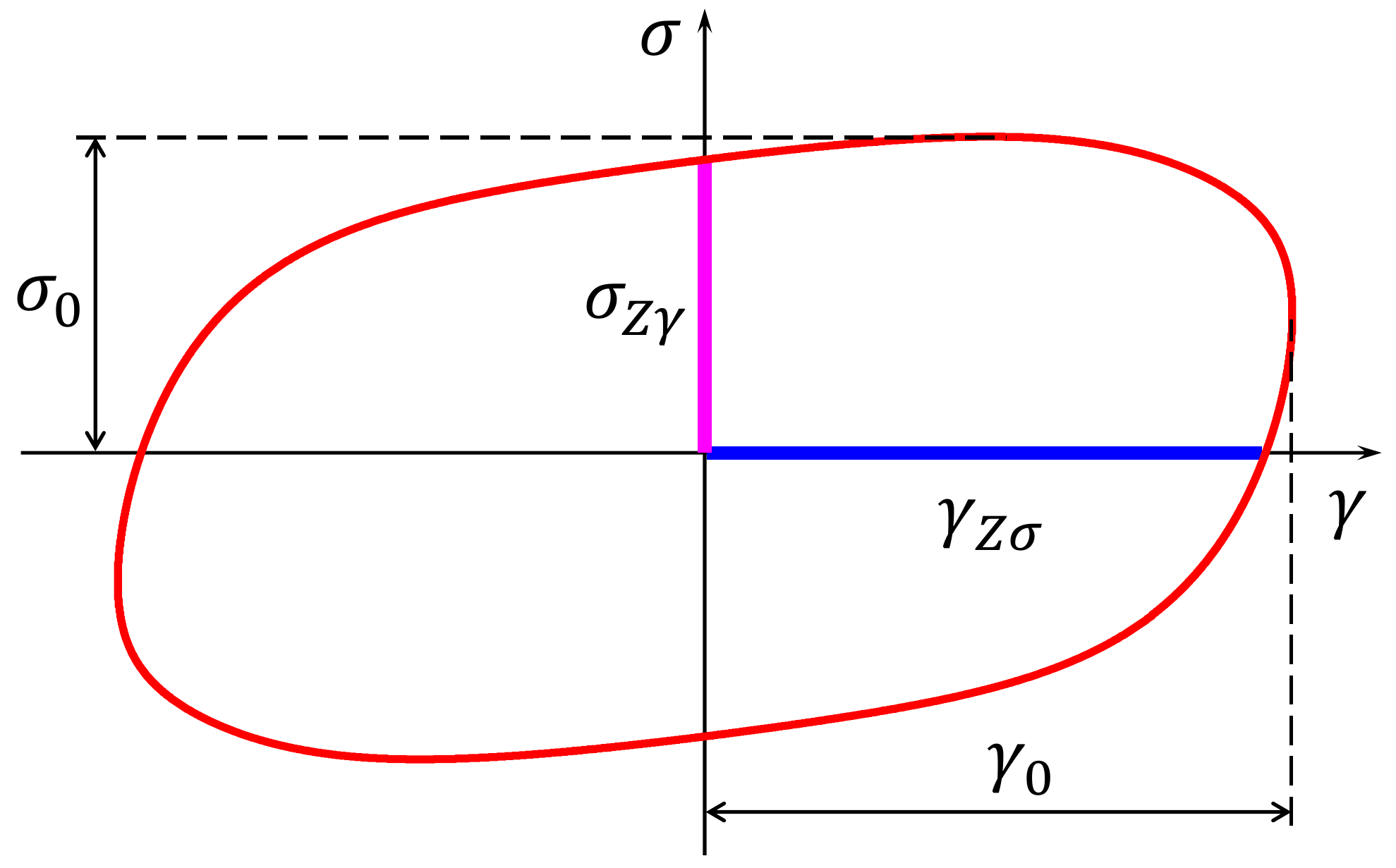}
\caption{Definition of the quantities $\gamma_{Z\sigma}$ and $\sigma_{Z\gamma}$.}
\label{fig:Fig5a}
\end{figure}

Finally, for the sake of completeness, let us introduce the zero-strain based generalized loss angle, $\delta_{Z\gamma}$, by the formula 
\begin{equation}
\delta_{Z\gamma}(\omega,\gamma_0)={\rm arcsin\,}\frac{\sigma_{Z\gamma}}{\sigma_0},
\label{1rh(4.5)}
\end{equation}
where $\sigma_{Z\gamma}$ is the stress at which the strain is instantaneously zero. Again, this quantity can be calculated from the Lissajous--Bowditch curve (see Fig.~\ref{fig:Fig5a}). 

For the sake of completeness, in the case of LAOS measurements represented via the stress versus strain rate Lissajous--Bowditch plot, we can define the zero-stress based, $\delta_{z\sigma}$, and the zero-strain-rate based, $\delta_{z\dot\gamma}$, generalized loss angles as follows:
\begin{equation}
\delta_{z\sigma}={\rm arccos\,}\frac{\sigma_{z\dot\gamma}}{\sigma_0},\quad
\delta_{z\dot\gamma}={\rm arccos\,}\frac{\dot\gamma_{z\sigma}}{\omega\gamma_0}.
\label{1rh(4.6)}
\end{equation}
Here, $\sigma_{z\dot\gamma}$ is the stress at which the strain rate is instantaneously zero, $\dot\gamma_{z\sigma}$ is the strain rate at which the stress is instantaneously zero.

\section{Discussion}
\label{secSL6}

In this section, we briefly overview the defined generalized moduli on an illustrative example of LAOS and formulate some recommendations and conclusions.

\subsection{Example of strain-controlled LAOS}
\label{secSL6A}

We consider a LAOS flow of a molten low-density polyethylene at $150^\circ$ with the test frequency $f=1$~Hz (so that $\omega=2\pi$ rad/s) and the strain amplitude $\gamma_0=10$, which was treated in the FT framework by \citet{GiacominOakley1993} using the discrete Fourier transform to obtain the Fourier series coefficients from the LAOS loop (see Table~\ref{tab:table1}).

\begin{table}[h]
\caption{FT coefficients (Pa) \cite{GiacominOakley1993} and SD coefficients (Pa).}
\begin{ruledtabular}
\begin{tabular}{crrrr}
$m$ & $G^\prime_m(\omega,\gamma_0)$ & $G^{\prime\prime}_m(\omega,\gamma_0)$ &
${\bf G}^\prime_m(\omega,\gamma_0)$ & ${\bf G}^{\prime\prime}_m(\omega,\gamma_0)$
\\
\colrule
1 & 2188{.}962 & 6681{.}102 & 1554{.}936 & 8276{.}825 \\
3 & -416{.}276 & -576{.}721 & -3{.}114 & -17{.}58 \\
5 & 166{.}008 & -22{.}007 & 0{.}148 & -0{.}022 \\
7 & -40{.}880 & 5{.}802 & -0{.}002 & -0{.}001 \\
9 & 7{.}880 & 1{.}801 & 0{.}00002 &0{.}000005
\end{tabular}
\end{ruledtabular}
\label{tab:table1}
\end{table}

The coefficients ${\bf G}^\prime_m$ and ${\bf G}^{\prime\prime}_m$ for the stress decomposition (\ref{1rh(1.4)}) (see Table~\ref{tab:table1}) were evaluated using formulas (\ref{1rh(1.10)}) and (\ref{1rh(1.11)}), which, in this case, are specified as follows:
$$
{\bf G}^\prime_1=G^\prime_1+3G^\prime_3+5G^\prime_5+7G^\prime_7+9G^\prime_9,
$$
$$
{\bf G}^{\prime\prime}_1=G^{\prime\prime}_1-3G^{\prime\prime}_3+5G^\prime_5
-7G^{\prime\prime}_7+9G^{\prime\prime}_9,
$$
$$
\gamma_0^2
{\bf G}^\prime_3=-4G^\prime_3-20G^\prime_5-56G^\prime_7-120G^\prime_9,
$$
$$
\gamma_0^2
{\bf G}^{\prime\prime}_3=4G^{\prime\prime}_3-20G^{\prime\prime}_5
+56G^{\prime\prime}_7-120G^{\prime\prime}_9,
$$
$$
\gamma_0^4
{\bf G}^\prime_5=16G^\prime_5+112G^\prime_7+432G^\prime_9,
$$
$$
\gamma_0^4
{\bf G}^{\prime\prime}_5=16G^{\prime\prime}_5-112G^{\prime\prime}_7
+432G^{\prime\prime}_9,
$$
$$
\gamma_0^6
{\bf G}^\prime_7=-64G^\prime_7-576G^\prime_9,
$$
$$
\gamma_0^6
{\bf G}^{\prime\prime}_7=64G^{\prime\prime}_7-576G^{\prime\prime}_9,
$$
$$
{\bf G}^\prime_9=\frac{256}{\gamma_0^8}G^\prime_9,\quad
{\bf G}^{\prime\prime}_9=\frac{256}{\gamma_0^8}G^{\prime\prime}_9.
$$

First, using formulas (\ref{1rh(2.8a)}) and (\ref{1rh(2.8b)}), we readily get the energetically equivalent storage and loss moduli
$$
G^\prime_E=2188{.}962~{\rm Pa}, \quad
G^{\prime\prime}_E=6681{.}102~{\rm Pa},
$$
which coincide with the FT coefficients ${\bf G}^\prime_1$ and ${\bf G}^{\prime\prime}_1$.

Now, formulas (\ref{1rh(3.5)}) and (\ref{1rh(3.18)}), we get the minimum strain storage and loss moduli
$$
G^\prime_M=1554{.}936~{\rm Pa}, \quad
G^{\prime\prime}_M=8276{.}825~{\rm Pa},
$$
which simply coincide with the SD coefficients $G^\prime_1$ and $G^{\prime\prime}_1$.

The so-called large strain storage and loss moduli can be evaluated suing formulas (\ref{1rh(3.4)}), (\ref{1rh(3.6)}) and (\ref{1rh(3.15)}), (\ref{1rh(3.16)}) as
$$
G^\prime_L=2820{.}006~{\rm Pa}, \quad
G^{\prime\prime}_L=6089{.}977~{\rm Pa}.
$$

Further, numerically evaluating the stress amplitude $\sigma_0=65745{.}8~{\rm Pa}$, by formula (\ref{1rh(4.1)}) we obtain the generalized dynamic modulus
$$
G_D=6574{.}58~{\rm Pa},
$$
whereas formulas (\ref{1rh(4.2)}), (\ref{1rh(4.4)}), and (\ref{1rh(4.5)}) yield three different loss angles (in radians)
$$
\delta_A=1{.}065,\quad
\delta_{Z\sigma}=1{.}277,\quad
\delta_{Z\gamma}=1{.}184.
$$

It is clear that some other quantities of interest can be evaluated in the same way. 

\subsection{Discussion and conclusion}
\label{secSL6B}

As it is shown by the example of the previous section, the discrepancy between the moduli evaluated in LAOS according to different definitions can be very substantial. For example, $G^\prime_E$ and $G^\prime_L$ differ from $G^\prime_M$ by 40\% and 80\%, respectively, whereas the three storage moduli must coincide in SAOS.

It is interesting to observe that the SD coefficients show tendency to decay faster than the FT coefficients (see Table~\ref{tab:table1}). Therefore, the fit of the experimental LAOS loop \citep{GiacominOakley1993} will require a smaller number of SD coefficients. 

Further, it is important to emphasize that though the quantities ${\cal G}^\prime_m$ and ${\cal G}^\prime_l$ were introduced in a similar way to $G^\prime_M$ and $G^\prime_L$, they do not reduce to the storage modulus in SAOS. However, the corresponding reciprocal values, $J^\prime_m$ and $J^\prime_l$ (see formulas (\ref{1rh(5.8a)}) and (\ref{1rh(5.8b)})) coincide with the storage compliance in the range of small strains. Therefore, it is of practical interest to compare the values of $J^\prime_m$ and $J^\prime_l$, which are determined under strain control, with the small-strain compliance, $J^\prime_M$, and the large-strain compliance, $J^\prime_L$, which are defined under stress control with the sinusoidal input as follows \citep{LaugerStettin2010}:
\begin{equation}
J^\prime_M(\omega,\sigma_0)=\frac{d\gamma}{d\sigma}\Bigr\vert_{\sigma=0},\quad
J^\prime_L(\omega,\sigma_0)=\frac{\gamma}{\sigma}\Bigr\vert_{\sigma=\pm\sigma_0}.
\label{1rh(6.1)}
\end{equation}
Note that comparison of strain-controlled and stress-controlled LAOS flows was performed in \citep{LaugerStettin2010,BaeLeeCho2013}.

Let us underline again that though the right-hand sides of two formulas (\ref{1rh(6.1)}) coincide with the right-hand sides of (\ref{1rh(5.8a)}) and (\ref{1rh(5.8b)}), respectively, in LAOS, the values of $J^\prime_M$ and $J^\prime_L$ can be different from $J^\prime_m$ and $J^\prime_l$, even for the same level of the stress amplitude. 

A very important finding in our study is the exact coincidence of the minimum-strain storage modulus $G^\prime_M$ and the minimum-rate loss modulus $G^{\prime\prime}_M$ with the first SD coefficients ${\bf G}^\prime_1(\omega)$ and ${\bf G}^{\prime\prime}_1(\omega)$, respectively. In the spirit of the decomposition method of \citet{ChoHyunAhn2005}, the latter quantities, being coefficients of the leading terms in the SD expansions (\ref{1rh(1.6)}) and (\ref{1rh(1.7)}), should not depend on the strain amplitude $\gamma_0$. This fact is of crucial relevance for LAOS, where other considered generalized moduli are strain amplitude dependent. 

Thus, to conclude, a rheologicaly nonlinear material in LAOS is characterized by a set of storage and loss moduli, each of which has a more or less distinct physical meaning. While the different definitions result in the same values in SAOS, their predictions in LAOS can differ by tens of percents (of course, depending on the level of the strain amplitude). Therefore, when characterizing a LAOS flow in terms of the storage and loss moduli provided by commercially available rheometers, a special attention should be paid to specifying the method of moduli measurement. The strain-amplitude independent generalized storage and loss moduli $G^\prime_M$ and $G^{\prime\prime}_M$ are highlighted as being worthy of identifying the mechanical properties of a rheologically nonlinear material in LAOS.

\begin{acknowledgments}
IA is grateful to the Biofilms center for the hospitality during his stay at the Malm\"o University, where this research was carried out.
\end{acknowledgments}

\end{document}